\documentclass[aps,prb,reprint,longbibliography]{revtex4-2}
\usepackage{graphicx}
\usepackage{amsmath}
\usepackage{amssymb}
\usepackage{hyperref}
\usepackage{gensymb}
\usepackage[version=4]{mhchem}
\begin{document}
\title{Polarization switching in sliding ferroelectrics: the roles of fluctuation and domain wall}
\author{Ziwen Wang}
\author{Shuai Dong}
\email{sdong@seu.edu.cn}
\affiliation{Key Laboratory of Quantum Materials and Devices of Ministry of Education, School of Physics, Southeast University, Nanjing 211189, China}
\date{\today}

\begin{abstract}
Sliding ferroelectricity is highly attractive for its low energy barriers and fatigue resistance. As the origin of these exotic properties, its unconventional switching dynamics remains poorly understood: how an electric field drives a perpendicular sliding? Taking $h$-BN bilayer as a model system, its switching dynamics is studied using \textit{ab initio} calculations. The off-diagonal Born effective charge leads to the perpendicular relationship between the electric field and ionic movements. Interestingly, the rules of intrinsic coercive field are distinct between $h$-BN bilayer and conventional ferroelectrics. For $h$-BN bilayer, any perturbation breaking the in-plane symmetry plays a key role to assist the avalanche-like switching dynamics. Moreover, the exotic large off-diagonal Born effective charge near the $P=0$ intermediate state results in a wriggling motion of domain walls in $h$-BN bilayer. Our results reveal the key factors in the ferroelectric switching of sliding ferroelectrics at room temperature.
\end{abstract}
\maketitle

{\it Introduction.}~Ferroelectrics are those materials with spontaneous electric polarizations, which are switchable under external electric field. These switchable polarizations have significant impacts for a wide range of technological applications, including nonvolatile memory, sensors, and photonic devices~\cite{Scott2007,Rabe2007}. In the pursuit of higher integration and device miniaturization, ferroelectrics with reduced scale to nanometer are highly demanded. Comparing with conventional ferroelectric films based on three-dimensional oxides, two-dimensional (2D) ferroelectrics have many inborn superiorities and thus have attracted great research enthusiasms in recent years \cite{Li2024,Wang2023:nm,Guan2020}. The confirmed members in this family include few-layer \ce{CuInP2S6} \cite{Belianinov2015,Liu2016}, monolayer SnTe \cite{Chang2016}, $\alpha$-\ce{In2Se3} nanoflakes \cite{Cui2018,Xiao2018,Zhou2017}, monolayer Bi \cite{Gou2023}, and so on.

Among these 2D ferroelectrics, a novel branch termed sliding ferroelectrics has been extensively attentioned, which exhibits an unconventional origin of polarizations \cite{Li2017,science2021-MBS,science2021-PJH,Wu2021}. Specifically, some stacking modes of van der Waals (vdW) layers break the inversion symmetry and thus generate out-of-plane electric dipoles. Then these dipoles can be modulated and even reversed by changing the stacking modes, typically via the interlayer sliding or twisting. Shortly after the theoretical prediction by Wu \textit{et al.} \cite{Li2017}, sliding ferroelectricity has been experimentally confirmed in several systems, including \ce{WTe2} bilayer/few-layer \cite{Fei2018,Xiao2020}, bilayer hexagonal BN ($h$-BN) \cite{science2021-MBS,science2021-PJH}/1T$^\prime$-\ce{ReS2} \cite{Wan2022}/3R-\ce{MoS2} \cite{science2024}, vdW heterobilayers such as \ce{MoS2}/\ce{WS2} \cite{rogee2022}, as well as an organic-inorganic hybrid crystal \cite{Miao2022}. Additionally, its unique switching behaviors have also been investigated, such as the ultrafast switching induced by light and possible polarization switching pathways with multiple interfaces \cite{PRL2024-MS,PRL2024-YRR,Liang2025}. 

Despite these advances, a fundamental question remains unresolved: how an out-of-plane electric field induces the in-plane sliding motion of atoms/ions? Such a perpendicular relationship of electrostatic property seems to mimic the Hall effect of electric current, implying plenty emergent physics hidden behind the phenomena. In fact, there are some paradoxes in sliding ferroelectrics. 
For instance, despite their tiny energy barriers, how can the sliding ferroelectricity persist to room temperature \cite{science2021-MBS,science2021-PJH,science2024}. Moreover, the recently observed high speed of ferroelectric switching remains incompletely understood \cite{Yasuda2024,He2024}. Thus, it is crucially important to understand the switching dynamics in sliding ferroelectrics.

In this Letter, the switching dynamics of sliding ferroelectricity is studied using density functional theory (DFT) calculations and \textit{ab initio} molecular dynamics (AIMD) simulations. Among all confirmed sliding ferroelectrics, $h$-BN owns the simplest chemical formula, as well as the largest polarization in this family \cite{Wu2021}. Thus, the $h$-BN bilayer will be studied as the model system to reveal the switching dynamics. Details of numerical methods can be found in Supplemental Material (SM)~\cite{sm}. Our calculations reveal that nonzero off-diagonal element of Born effective charge emerges during the ferroelectric switching process, despite its zero value in the initial ferroelectric state. The required critical electric field shows a progressive reduction accompanying the sliding, implying an avalanche-like behavior, which is different to conventional ferroelectrics like \ce{BaTiO3} where an intrinsic maximum field appears during the polarization reversal. Furthermore, other effects, e.g. wrinkle, layer gap, and temperature, have been investigated, which can modulate the coercive field differentially. Notably, the large off-diagonal Born effective charge near the $P=0$ intermediate state leads to a wriggling sliding motion of domain walls in multi-domain $h$-BN bilayer, which may be responsible to the high speed of the polarization switching.

\begin{figure*}
	\includegraphics[width=\textwidth]{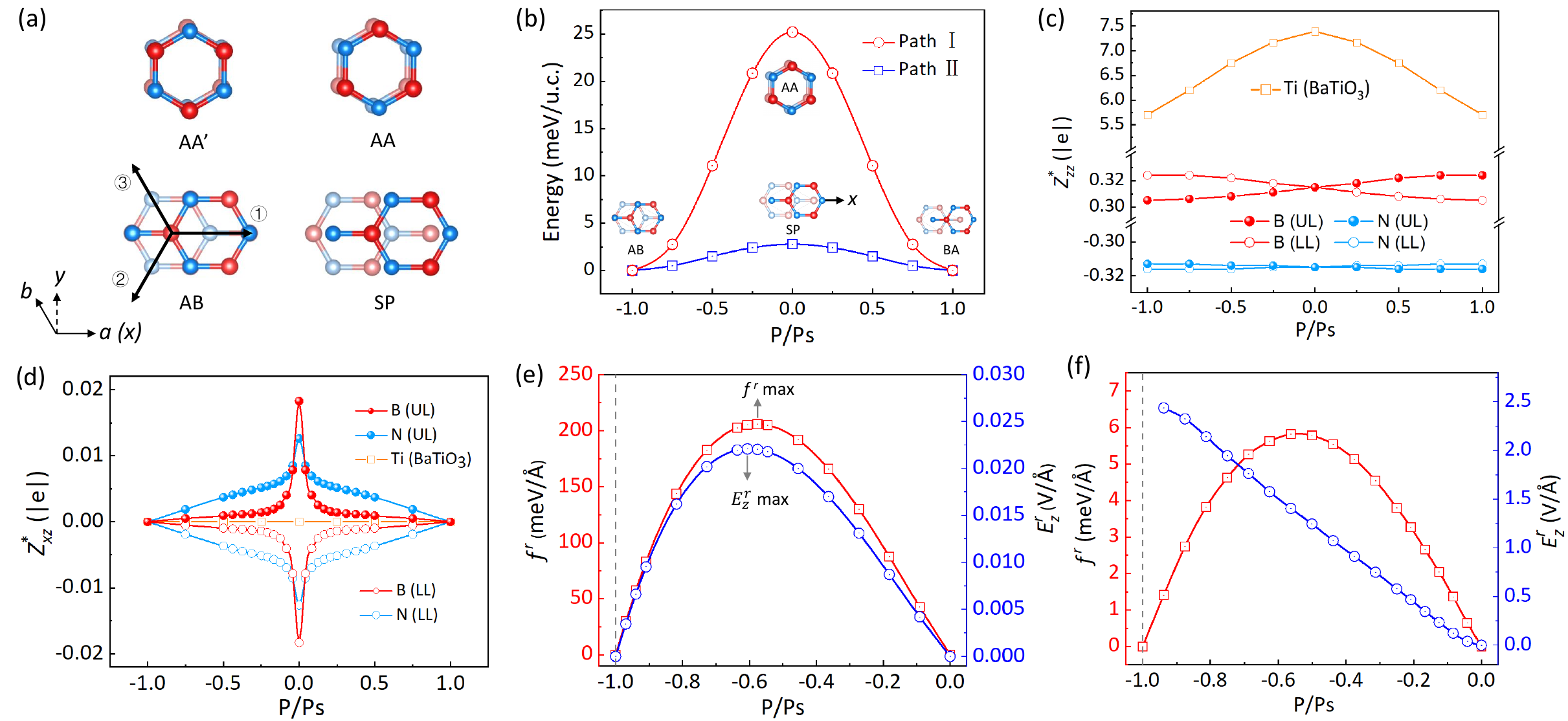}
	\caption{(a) Top views of AA$^\prime$, AA, AB, and SP (saddle point) stacking modes of $h$-BN bilayer. The in-plane $C_3$ symmetry is denoted by black arrows in the AB mode. (b) Calculated energy barriers of two switching pathways of $h$-BN bilayer. Inset: Four typical configurations during the switching. AB: $+P$, AA or SP: $P=0$, BA: $-P$. (c-d) Tensor elements of Born effective charge in $h$-BN bilayer, in comparison with \ce{BaTiO3}. (c) Diagonal element $Z^*_{zz}$. (d) Off-diagonal element $Z^*_{xz}$. UL: Upper layer. LL: Lower layer. (e-f) The intrinsic resistance force $f^r$ and required critical electric field $E_z^r$ during the polarization switching: (e) \ce{BaTiO3}. (f) $h$-BN bilayer.}
	\label{F1}
\end{figure*}

{\it Results \& discussion.}~Several possible high-symmetry stacking modes have been proposed for bilayer $h$-BN \cite{prl2013}, as illustrated in Fig.~\ref{F1}(a). Among these, the centrosymmetric AA$^\prime$ mode (point group $\bar{3}m$) is the most stable one, while the AB (or BA) mode (polar point group $3m$) is the first metastable one. Both AA$^\prime$ and AB have been experimentally observed in bilayer $h$-BN \cite{Acsnano2010,science2021-MBS}. Noting that the AA$^\prime$ and AB modes can not be interconverted via the interlayer sliding operation, but can be interconverted via the $60^\circ$ interlayer twisting operation plus the sliding operation.

A spontaneous polarization is expected in the AB (BA) stacking mode, which originates from the asymmetric vertical alignment of N and B atoms of different layers. The distortion of the $p_z$ orbital of N ion results in a net out-of-plane electric dipole \cite{science2021-PJH}. The degenerate AB and BA stacking can be interconverted through interlayer sliding of BN layers, leading to the reversal of ferroelectric polarization. Two distinct ferroelectric switching pathways have been examined: AB$\rightarrow$AA$\rightarrow$BA (Path I) and AB$\rightarrow$SP$\rightarrow$BA (Path II). The energy barrier for Path II is much lower than that of Path I [Fig.~\ref{F1}(b)], consistent with previous studies \cite{PRL2024-MS,PRL2024-YRR}. Therefore, only the Path II will be considered in the following.

\begin{figure*}
	\includegraphics[width=\textwidth]{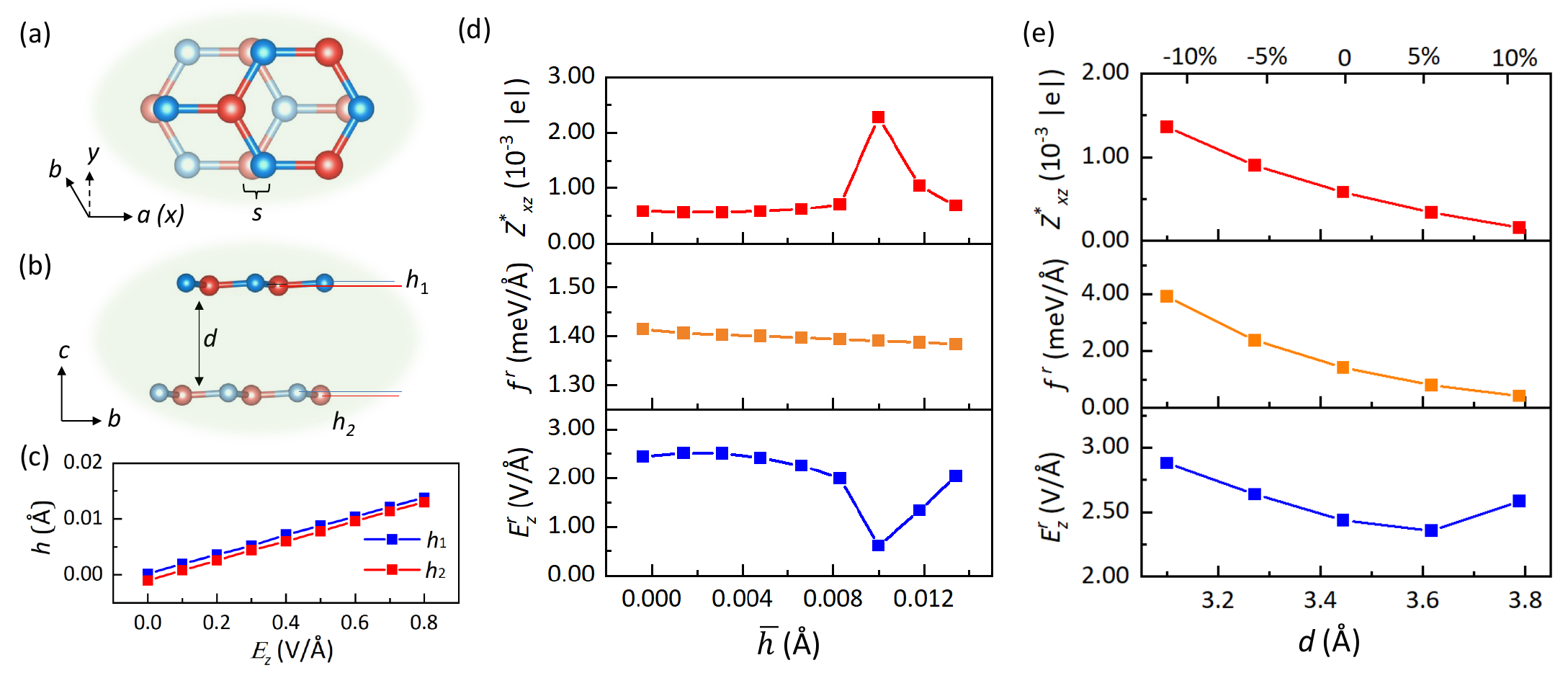}
	\caption {(a) Top view of the tiny sliding $s$ along the $x$-direction from the original AB stacking one. (b) Side view of the wrinkled structure. $h_1$ and $h_2$: the wrinkled heights of the upper and lower layer, respectively. $d$: the layer gap. (c) The evolution of wrinkled amplitudes as a function of vertical electric field $E_z$. (d-e) Effects of the layer wrinkle $\bar{h}$ and layer gap $d$. Upper: the sum of off-diagonal $Z^*_{xz}$ of B and N ions of the upper layer. The $Z^*_{xz}$ of lower layer is exactly opposite (not shown here). Middle: the intrinsic resistance force $f^r$ of one layer. Lower: the estimated required critical electric field $E_z^r$. (d) $\bar{h}$-dependent properties at the optimal $d=3.44$ \r{A}. (e) $d$-dependent properties without wrinkle.}
	\label{F2}
\end{figure*}

As stated before, it is not so intuitive how an electric field drives collective motion of ions perpendicularly. This paradox can be understood within the framework of Born effective charge \cite{PRB1997,Ghosez2000}. The Born effective charge $Z^*$ is a tensor: 
\begin{equation}
	Z^{*}_{\kappa,ij}=\Omega\frac{\partial P_j}{\partial u_{\kappa,i}}=-\frac{\partial ^2  F}{\partial E_j\partial u_{\kappa,i}}=\frac{\partial f^E_{\kappa,i}}{\partial E_j}, 
	\label{eq1}
\end{equation}
where $\kappa$ is the index of ions; $i$ and $j$ $\in\{x, y, z\}$ are three orthogonal orientations. $P$, $u$, $F$, $E$, and $\Omega$ denote the polarization, ion position, electric enthalpy, external electric field, and unit-cell volume, respectively. To drive the ion motion, the electrostatic force $f^E$ on the $\kappa$-ion under electric field $E$ can be expressed as:
\begin{equation}
	f^E_{\kappa,i}=\sum_j E_jZ^{*}_{\kappa,ij}. 
	\label{eq2}
\end{equation}
During the switching, the external electric field should be large enough to overcome the intrinsic resistance from energy barrier, which can be expressed as $f^r_{\kappa,i}=-\partial H/\partial u_{k,i}$ where $H$ is the Hamiltonian. The coercive field $E^c$ is defined as the maximum value of the required critical electric field $E^r$ during the polarization reversal.

Taking the conventional ferroelectric as an example, e.g. tetragonal perovskite \ce{BaTiO3}, its Born effective charge tensor $Z^*$ is diagonal, while all off-diagonal elements are zero~\cite{PRB1998}. During the ferroelectric switching between $\pm P$ (along the $z$-axis), the related $Z^*_{zz}$ obtained in our density functional perturbation theory (DFPT) calculation of the core ion Ti is shown in Fig.~\ref{F1}(c), which is rather large and evolves moderately as a function of $P$. The complete Born effective charge tensor for barium and oxygen ions are provided in Fig.~S1 of SM~\cite{sm}. Then the intrinsic resistance force $f^r$ (on positive ions) during the switching is estimated in Fig.~\ref{F1}(e), which starts from zero and reaches its maximum in the middle, corresponding to the steepest slop of the energy barrier (Fig.~S2 of SM~\cite{sm}). The required critical electric field $E_z^r$ to overcome such resistance force is derived from Eq.~\ref{eq2}, as shown in Fig.~\ref{F1}(e), which shows a similar tendency to $f^r$. The moderate $P$-dependence of $Z^*_{zz}$ leads to slight difference between the peaks of $f^r$ and $E_z^r$. In short, the $E_z^r$ for \ce{BaTiO3}'s (single-domain) switching is a climbing-like behavior with the maximum appearing around the peak of resistance force.

\begin{figure*}
	\includegraphics[width=\textwidth]{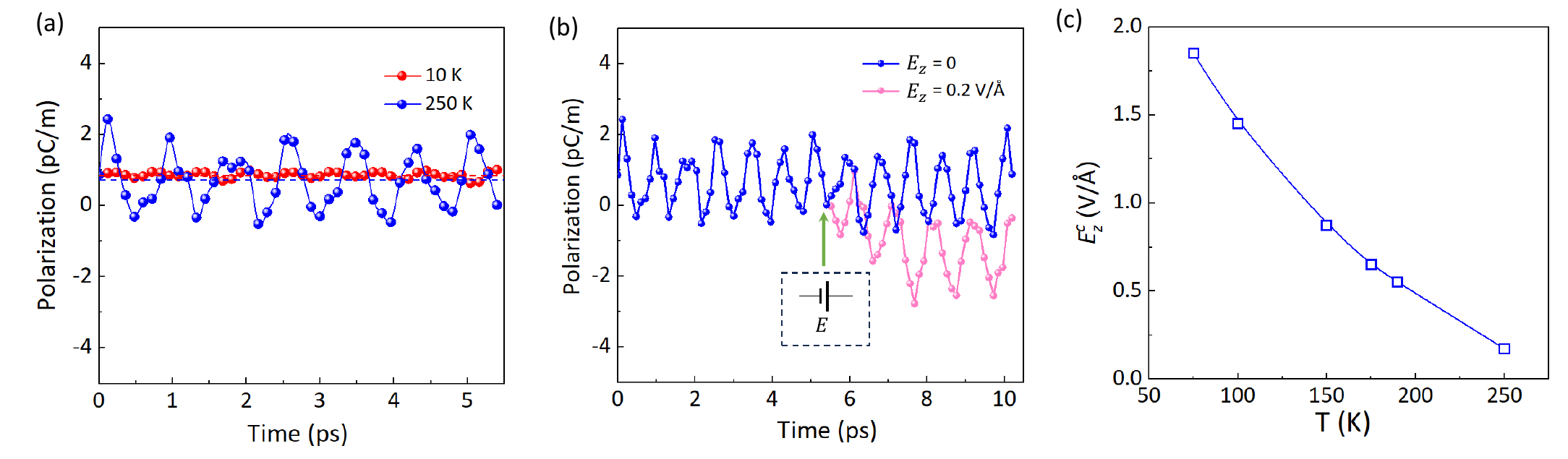}
	\caption{AIMD simulations of $h$-BN bilayer. (a) Time dependent polarization under different temperatures without the electric field. The average values of polarization are depicted in dotted lines. (b) Time dependent polarization under $250$ K with/without the electric field ($E_z$ = 0.2 V/\AA). The electric field is turn on at $5.4$ ps. (c) Temperature dependence of the magnitude of coercive field $E_z^c$ estimated from AIMD simulation. }
	\label{F3}
\end{figure*}

In contrast, the $Z^*_{zz}$ of B and N ions in $h$-BN bilayer are much smaller, implying weaker ionic behavior. More importantly, nonzero off-diagonal $Z^*_{xz}$ appears (except for the initial and end states), as shown in Fig.~\ref{F1}(d), which is responsible for the perpendicular relationship between electric field and sliding motion. The $Z^*_{xz}$'s of N and B ions of one layer own the same sign, implying the same-direction motion under a perpendicular electric field, i.e. the horizontal sliding instead of wrinkling. The values of $Z^*_{xz}$'s are much smaller than their corresponding $Z^*_{zz}$'s. An important fact is that $Z^*_{xz}$'s change drastically during the switching process, leading to a sharp peak at the $P=0$ state (i.e. the SP mode). The zero-value of $Z^*_{xz}$'s at the initial and end states (i.e. the AB and BA modes) is due to the in-plane $C_3$ rotational symmetry of the hexagonal lattice. In other words, there are three equivalent sliding paths with $120^\circ$ interval for the AB$\rightarrow$SP$\rightarrow$BA switching. Then an out-of-plane electric field can not break the $C_3$ rotational symmetry, and thus can not determine which one of the three is preferred. Thus the $Z^*_{xz}$ at the AB and BA modes should be exact zero.

These characteristics of $Z^*_{xz}$'s lead to a nontrivial switching process for $h$-BN bilayer, as shown in Fig.~\ref{F1}(f). Although the tendency of force $f^r$ looks similar to that of \ce{BaTiO3} which starts from zero and reaches the maximum in the middle, the required critical electric field $E_z^r$ exhibits totally different behaviors. At the initial ferroelectric state (e.g. the AB mode), the zero $Z^*_{xz}$'s make the system stable against the out-of-plane electric field. Only after a finite sliding, $E_z^r$ becomes finite and monotonously decreases with sliding (e.g. decreasing $P$). Thus, the largest $E_z^r$ is not an intrinsic constant, but determined by the initial sliding conditions, which can be affected by many extrinsic factors that break the initial  $C_3$ symmetry, such as thermal fluctuations, twist angles, as well as domain walls (to be discussed later). In short, the behavior of $E_z^r$ for sliding switching is avalanche-like, in contrast to conventional climbing-like behavior. This mechanism also applies to other sliding ferroelectrics with $C_3$ rotational symmetry, such as 3R-\ce{MoS2} bilayer and WS$_2$/MoS$_2$ heterobilayer, as shown in Figs.S3-S6 of SM \cite{sm}.

Beginning from a staggered stacking mode with tiny sliding $s=0.045$~\r{A} [Fig.~\ref{F2}(a)], the calculated vertical coercive field $E_z^c$ is $2.43$~V/\r{A}. Such a coercive field $E_z^c$ is higher than the experimental measured values ($\sim0.03$ V/\r{A}) for nearly two orders of magnitude \cite{science2021-MBS}. Thus, other factors should be considered to give a more realistic description.

Intuitionally, the vertical wrinkling of $h$-BN will be induced by the out-of-plane electric field. The heterocharge $Z^*_{zz}$'s of B and N ions [Fig.~\ref{F1}(c)] lead to opposite displacements along the $z$-axis under the vertical electric field $E_z$. The structure is optimized under the vertical electric field, and the layer thickness $h$ is defined to quantitatively characterize this wrinkling effect [Fig.~\ref{F2}(b)]. As shown in Fig.~\ref{F2}(c), $h$ grows linearly with increasing $E_z$, and the wrinkling is almost identical between two layers ($h_1$ is only slightly larger than $h_2$ due to the existence of polarization). However, the in-plane $C_{3}$ symmetry remains preserved, despite the wrinkling. Therefore, the avalanche-like behavior of $E_z^r$ will not be qualitatively altered by the wrinkling. Then the physical properties with wrinkling are recalculated by starting from a slightly staggered configuration $s=0.045 $ \r{A}, as summarized in Fig.~\ref{F2}(d). With increasing wrinkling amplitude [characterized using $\bar{h}=(h_1+h_2)/2$], the $Z^*_{xz}$ is almost a constant in the beginning, but owns a peak at $\bar{h}=0.01$ \r{A}. Meanwhile, the intrinsic resistance force $f^r$ on each layer is almost a constant. Consequently, $E_z^r$ will be significantly reduced at the peak of $ Z^*_{xz} $. For example, $ E_z^r\sim0.61$ V/\r{A} for $s=0.045$ \r{A} at $\bar{h}=0.01$ \r{A}, only $\sim1/4$ of the original one without wrinkling. In short, the vertical wrinkling, induced by the perpendicular electric field, can significantly lower $E_z^r$ only at a specific amplitude of vertical wrinkling.

Besides the vertical wrinkling, another tunable structural degree of freedom is the interlayer gap $d$, which is directly related to the sliding ferroelectricity~\cite{prb:wang, Ding2021}. Also taking the staggered $s=0.045$ \r{A} (and without the vertical wrinkling), the off-diagonal $Z^*_{xz}$ and $f^r$ are recalculated as a function of interlayer gap $d$. As shown in Fig.~\ref{F2}(e), both $Z^*_{xz}$ and $f^r$ change monotonously: the larger $d$, the smaller $Z^*_{xz}$ and $f^r$. Then the $E_z^r$, as a result of $f^r/Z^*_{xz}$, is not monotonous, which reaches a minimum at $d=3.62$ \r{A} (slightly larger than the original $d$). The minimum $E_z^r\sim1.90$ V/\r{A} is only $22\%$ lower than the original one. 

Above study has demonstrated that an initial nonzero $s$ is mandatory for sliding dynamics, while the vertical wrinkling and interlayer gap can affect this process. In real situations at finite temperatures, the comprehensive effect can be simulated using AIMD. As shown in Fig.~\ref{F3}(a), at $10$ K without external fields, the AB stacking mode is stable, with a ferroelectric $P_z\sim0.82$ pC/m. By increasing temperature to $250$ K, the thermal fluctuation is enhanced while an average $P_z\sim0.78$ pC/m is maintained. When an out-of-plane electric field $E_z=0.2$ V/\r{A} is applied at $250$ K, the polarization can be easily switched within a short time, as shown in Fig.~\ref{F3}(b). The required amplitudes of the coercive field $E_z^c$ at different temperatures obtained in AIMD are summarized in Fig.~\ref{F3}(c), which steeply decreases with increasing temperature. It is reasonable since the strong thermal fluctuations at high temperatures can trigger the avalanche-like switching more easily. Thus, the experimental coercive field at room temperature should be greatly suppressed, due to large $s$ from strong thermal fluctuations instead of the intuitional scenario of low energy barrier.

Finally, it is worth to investigate the effect of domain wall on the sliding dynamics. Previous experiments have observed the high speed of domain wall motion in the polarization switching process.\cite{science2024,Yasuda2024}. Given the largest off-diagonal $Z^*_{xz}$ in the symmetric intermediate SP state, the center region of domain wall is naturally expected to exhibit a much larger $Z^*_{xz}$ than that within flanks. Then the center region will be more sensitive to external field and thus plays as the pioneer of sliding dynamics. 

\begin{figure}
	\includegraphics[width=0.48\textwidth]{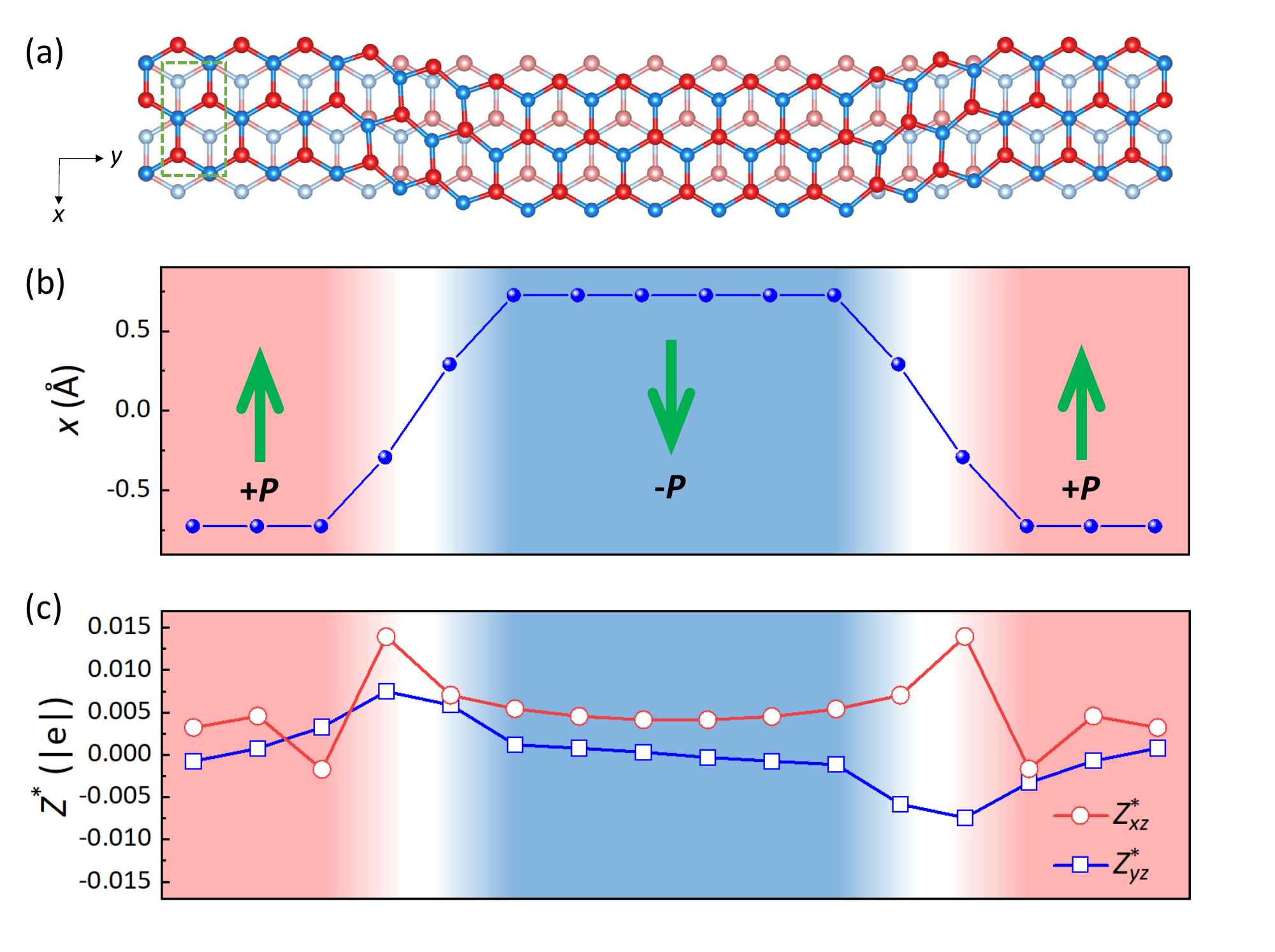}
	\caption{(a) Sketch of domain walls between two ferroelectric domains. Here the orthorhombic cell is taken as the unit cell, as depicted by the green rectangle. (b) The sliding distance of each unit cell ($x$) over the ferroelectric domains and domain walls. The domain regions with positive and negative polarization are denoted in green arrows and different background colors. The gradient background color regions denote the domain walls. (c) The off-diagonal Born effective charge elements $Z^*_{xz}$ and $Z^*_{yz}$ of each unit cell in the upper layer. The $Z^*_{xz}$ and $Z^*_{yz}$ become larger at center regions of domain walls.}
	\label{F4}
\end{figure}

To simulate the domain wall, a supercell containing $+P$, $-P$, and domain wall is constructed, as shown in Fig.~\ref{F4}(a). By fixing the $+P$ and $-P$ domains and relaxing domain walls, the profile of local $P$, characterized by the sliding distance $x$, is shown in Fig.~\ref{F4}(b). The sliding distance $x$ changes continuously from $+P$ to $-P$ states. Then elements of the off-diagonal Born effective charge $Z^*$ for each unit cell are calculated. As shown in Fig.~\ref{F4}(c), the center regions of domain walls own relative larger $Z^*_{xz}$ values, and even within the domains $Z^*_{xz}$'s become non-zero due to the proximity effect of domain walls. Furthermore, the off-diagonal $Z^*_{yz}$ element is also induced which further enhances the sliding. Thus, the center region of domain wall will achieve a much larger accelerated velocity and move faster than the flank regions when a vertical electric field is applied. This out-of-sync acceleration will generate wriggling and the faster-moving center region will pull the non-center regions, further accelerate the domain wall motion. In the subsequent moment, the newly formed center region induces the wriggling motion continuously and increases the domain wall velocity again until the complete ferroelectric switching process is finished. This characteristic may be responsible for the high speed of the ferroelectric switching observed in experiments, warranting further investigation.

In summary, the switching dynamics of sliding ferroelectric $h$-BN bilayer has been investigated numerically, which is nontrivially different from conventional ferroelectrics. It is the nonzero off-diagonal element of Born effective charge leading to perpendicular relationship between electric field and sliding motion. To start the sliding, perturbation induced by extrinsic factors such as thermal fluctuation, is needed to break the $C_3$ symmetry of the initial ferroelectric state. The required critical electric field $E_z^r$ of a single domain does not have an intrinsic maximum, conceptually different from conventional ferroelectrics like \ce{BaTiO3}. Layer wrinkle, layer gap, temperature, as well as domain wall can be helpful to reduce the ferroelectric coercive field.

\textit{Note:} We became aware of two recent preprints on the switching dynamics of sliding ferroelectrics~\cite{ke2025,shi2025}. In ref.~\cite{ke2025}, their work also highlighted the key effects of off-diagonal Born effective charges and domain walls. In ref.~\cite{shi2025}, an undamped soliton-like domain wall motion was reported~\cite{shi2025}.

\begin{acknowledgments}
 We thank Dr. Peng Chen and Dr. Junjie Zhang for helpful discussions. Work was supported by National Natural Science Foundation of China (Grant Nos. 12325401, 12274069, and 123B2053) and the Big Data Computing Center of Southeast University.
\end{acknowledgments}

\bibliography{h-BN}
\end{document}